\documentclass{PoS}
\usepackage{epsfig}
\usepackage{epstopdf}
\title{The ARCADE Raman Lidar and atmospheric simulations for the Cherenkov Telescope Array}

\ShortTitle{ARCADE and Simulations for the CTA}

\author{\speaker { L. Valore}$^a$, C. Aramo$^a$,
B.M. Dinelli$^b$, F. Di Pierro$^c$, G. Dughera$^c$, M. Gaug$^d$, M. Iarlori$^e$,
M. Marengo$^c$, E. Papandrea$^{b,f}$, E. Pietropaolo$^e$, V. Rizi$^e$, P. Vallania$^{c,g}$, C. Vigorito$^{c,h}$
for the CTA Consortium \\ 
\llap{$^a$} Universit\`a degli Studi di Napoli Federico II and INFN Napoli, Italy\\
\llap{$^b$} ISAC-CNR, Italy \\
\llap{$^c$} INFN Torino, Italy\\
\llap{$^d$} Unitat de F\'isica de les Radiacions, Universitat Aut\`onoma de Barcelona\\
\llap{$^e$} CETEMPS/DSFC Universit\`a degli Studi dell'Aquila, Italy\\
\llap{$^f$} Serco s.p.a.\\
\llap{$^g$} INAF Torino, Italy\\
\llap{$^h$}Dipartimento di Fisica dell'Universit\`a di Torino\\
E-mail: \email{laura.valore@na.infn.it}
}

\abstract{The Cherenkov Telescope Array (CTA) is the next generation of ground-based very high energy
gamma-ray Imaging Atmospheric Cherenkov Telescopes (IACTs). Since observations with this
technique are affected by atmospheric conditions, an accurate knowledge of the atmospheric
properties is fundamental to improving the precision and duty cycle of the CTA. Measurements of
absorption and scattering properties of the atmosphere, due to aerosols and molecules, can be
used either in the event reconstruction software or in a detailed atmospheric radiative transfer
model such as MODTRAN, an analytical code designed to model the propagation of
electromagnetic radiation. The output of the MODTRAN software is then used as an input for the
air shower simulation and Cherenkov light production, giving the optical depth profiles that together
with the refractive index allow the proper simulation of the gamma-ray induced signals and
a correct measurement of the primary energy from the detected
signals. The ARCADE Raman Lidar is part of the CTA baseline for the on-site characterization of
the aerosol attenuation profiles of the UV light. The collected data will be used in preparation for
the full operation of the array, providing nightly information about the aerosol properties on site 
at 355 nm, such as the vertical aerosol optical depth and the water vapour mixing ratio with an
altitude resolution better than 100 m from about 400 m to 10 km above ground level. These
measurements will help to define the needs for Monte Carlo simulations of the shower
development and of the detector response. This instrument will also be used for the intercalibration
of the future Raman Lidars that are expected to operate at the CTA sites. This contribution
includes a description of the ARCADE Lidar, that includes two Raman channels (nitrogen and
water vapour) and one elastic channel, and the characterization of the performance of the system
and test results in L'~Aquila. The system is expected to be shipped to the northern site of the CTA
(La Palma) before the end of 2017, to acquire data locally for 1 year before being moved to the
southern site (Chile).}

\FullConference{35th International Cosmic Ray Conference --- ICRC2017\\
		10--20 July, 2017\\
		Bexco, Busan, Korea}

\begin{document}

\section{Introduction}
The Cherenkov Telescope Array (CTA) is a new huge observatory for the detection at ground of very high energy (VHE)
gamma rays, based on more than 100 Imaging Air Cherenkov Telescopes (IACTs), to study the most energetic phenomena 
in the Universe with an unprecedented accuracy.
The CTA target is ambitious: the sensitivity with respect to present ground-based VHE observatories 
(H.E.S.S., MAGIC and VERITAS) will be improved by about one order of magnitude and the covered energy range 
will span from a few tens of GeV to above 300 TeV enhancing at the same time angular and energy resolution. 
The CTA Observatory will consist of two separate arrays,
one located in the Northern Emisphere (Observatorio Roque de Los Muchacos - ORM, La Palma, Spain) and one in the 
Southern Emisphere (Cerro Armazones, close to Paranal, Chile), to 
ensure full sky coverage. Each array will be composed of several tens of IACTs of different sizes 
(Small Size Telescope - SST, Medium Size Telescope - MST, Large Size Telescope - LST) that are designed to operate in 
different energy ranges. Working prototypes exist or 
are under construction for all the telescope designs. Construction of the first LST in La Palma is in progress.

 \begin{figure}[h!t]
   \centering
   \includegraphics[width=.9\textwidth]{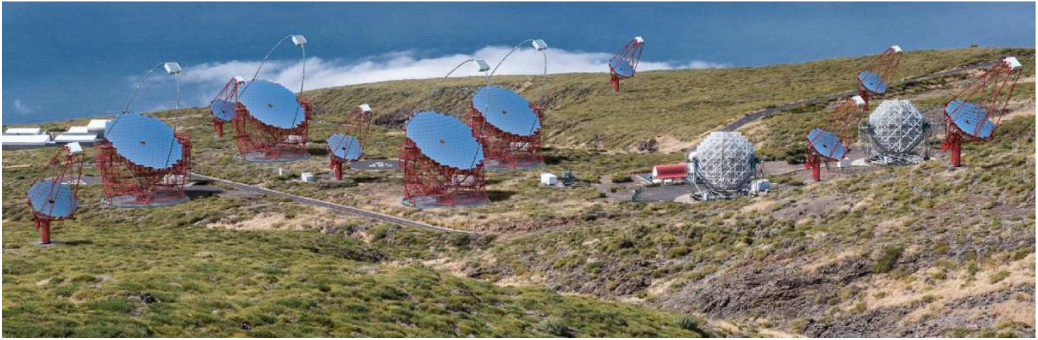}
  \caption{Rendering of the CTA northern array in La Palma}
  \label{CTA}
 \end{figure}

The development of the Extensive Air Showers generated by gamma primaries is greatly influenced by the atmosphere.
The variable atmospheric properties affect the observations of the Cherenkov light in multiple ways,  contributing
significantly to the systematic uncertainty on the primary energy and flux determination. 
The atmosphere plays a double role, being responsible at the same time for the production of the Cherenkov light and also for its 
attenuation when it travels towards the telescopes. In particular, the production of  Cherenkov light depends 
on the molecular profile, while the rapidly changing aerosol profile affects its attenuation.   
In the CTA, high quality of the data will be ensured by a constant monitoring of the local optical properties 
of the atmosphere during data taking with the aim to correct the collected data, as can be found elsewhere in these
proceedings \cite{jan_icrc17}. Both on-site atmospheric calibration instruments and external data (GDAS) will be 
used to monitor the atmospheric properties. The CTA sites will have a multi-instrument atmospheric monitoring system, 
that will include, among others, Raman Lidars that will provide measurements of the aerosol stratification in atmosphere during 
data taking and along the line of sight of the telescopes \cite{raman_icrc17}. The ARCADE Raman Lidar has a different target: it will be operating 
on each CTA site during the preparation of the array to provide a first survey of the local aerosol properties, that are needed to tune the 
simulations. The absorption and scattering properties of the atmosphere are input parameters in 
the atmospheric radiative transfer model MODTRAN, a commercial analytical code that describes the propagation of the electromagnetic radiation, which is used for the Monte Carlo simulation (which additionally includes the shower and detector response simulation) thus allowing to estimate the effect 
of the different atmospheric molecular and aerosol profiles on the CTA performance. In the following, the upgraded ARCADE Raman Lidar
and the Monte Carlo simulation studies based on various atmospheric scenarios are described.

\section{The ARCADE Raman Lidar}

The ARCADE Lidar Raman has been recently upgraded from a former telescope to become part of the CTA. The system was designed and
built in Italy within the Atmospheric Research for Climate and Astroparticle DEtection project \cite{arcade_icrc15} from 2012 to 2015, 
and took data for one year in Lamar, Colorado (USA) in a difficult environment, proving to be a reliable system.   
This instrument will be installed first at the northern site (La Palma) for a period of one to two years, then it is expected to move
to the southern site (Chile) to provide information on the aerosol properties during the construction of both arrays. 
It is not expected to operate during the future CTA data taking: other Raman Lidars are specifically designed for this purpose, 
as described in \cite{raman_icrc17}. The nightly and seasonal aerosol attenuation profiles of the UV light 
measured on site by ARCADE, whose laser source operates at 355 nm, will help to tune the atmospheric conditions 
in the Monte Carlo simulation chain of the shower development and of the detector response. 
Also, it will be used as a benchmark for the intercalibration of the Raman Lidars that will operate on the CTA sites.  
The data collected during the year of operation on site will be analyzed to provide nightly vertical profiles of the aerosol extinction 
coefficient or aerosol optical depth, nightly vertical profiles of aerosol backscatter coefficient, nightly vertical profiles of the water 
vapour mixing ratio. All of these parameters will be measured with a time resolution of about 5 to 15 minutes, and an altitude resolution below 100 m 
from about 400 m to 10000 m above ground level. In addition, we'll perform cross-correlations with measurements from the MAGIC LIDAR system working with a laser source 
at 532 nm (North site) and cross-correlation with the Sun/Moon photometer measurements \cite{Jurysek}, (South site).
Concerning the ORM site, measurements will be arranged in order to avoid any interference with the telescopes
already present on site. In the most conservative configuration, the data taking can be done 2 times per night for 
15 minutes at the beginning and at the end of the night, right before and after astronomical twilight, to minimize interference. This
approach is being presently used in the Pierre Auger Collaboration to run the Raman Lidar installed in the middle of the Observatory, and
data are successfully provided since 2013 \cite{raman_Auger}. 
Another option may be to add one set of measurements during the night for 15 minutes, when possibly affected telescopes point in 
other directions than the ARCADE Lidar. The group is collaborating with local institutions to define the best strategy to run our 
measurements and to define a suitable location.

\section{The instrument upgrade}

\begin{figure}[h!t]
   \centering
   \includegraphics[width=.43\textwidth]{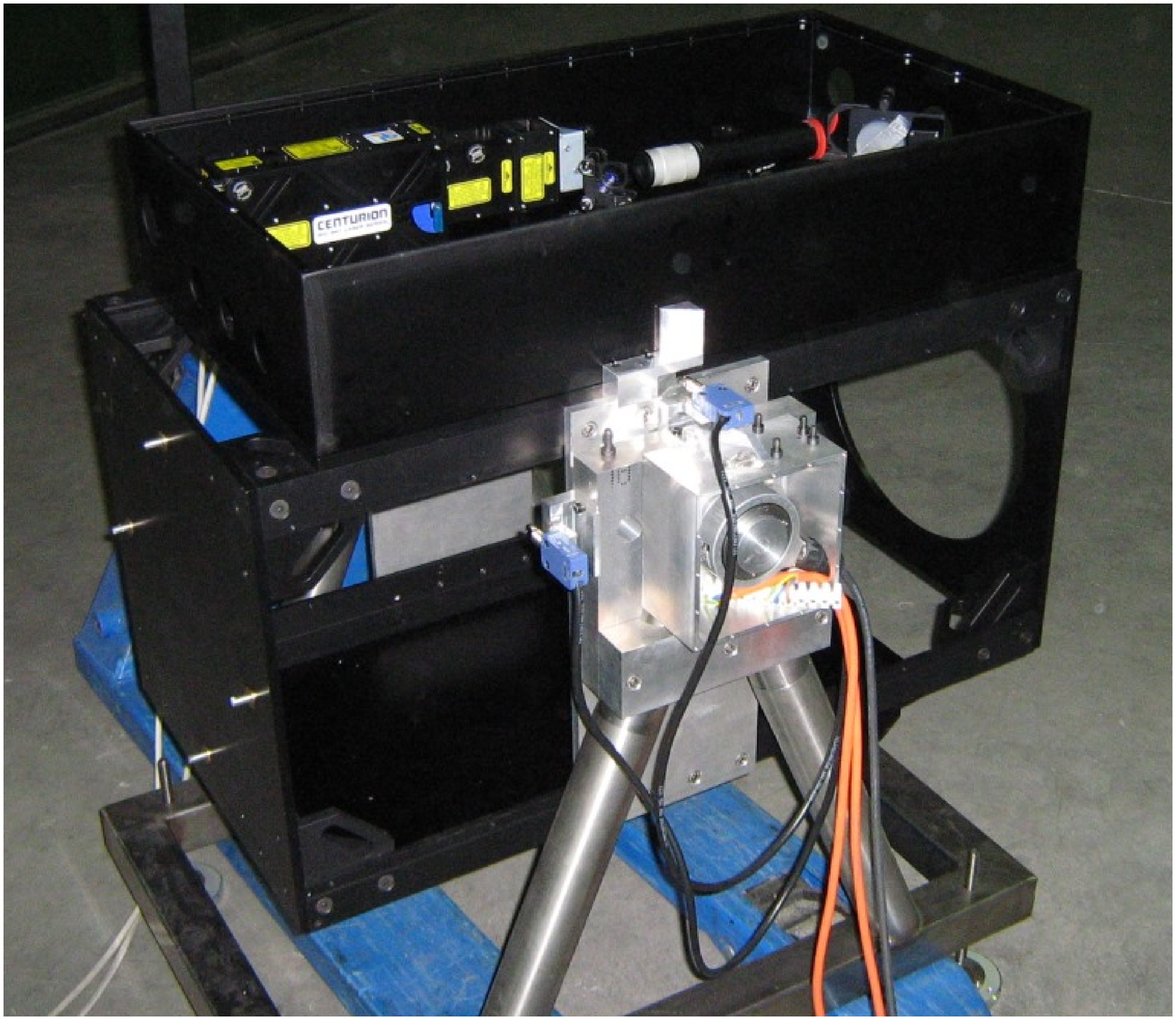}
   \includegraphics[width=.4\textwidth]{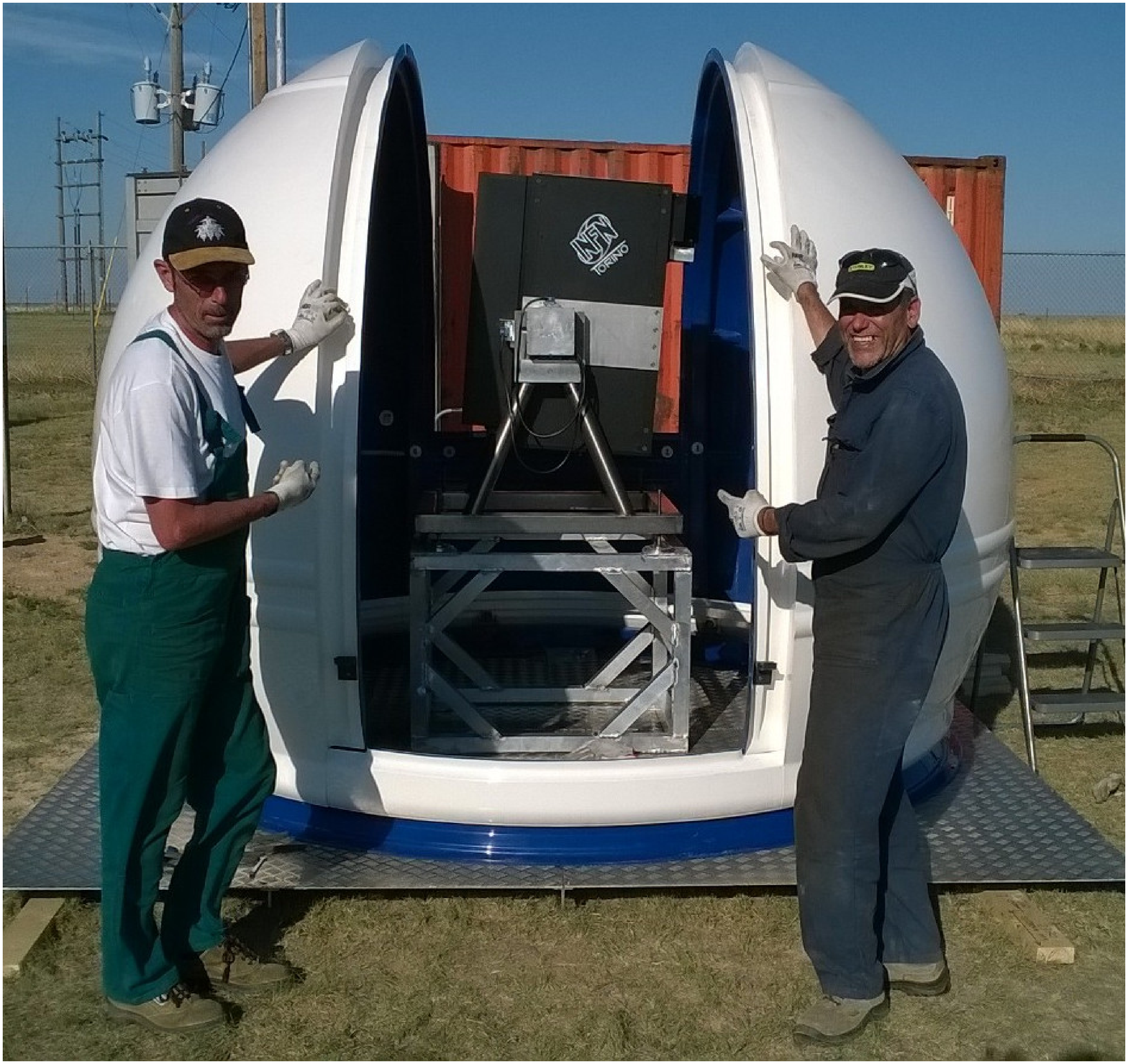}
  \caption{On the left, the ARCADE telescope. On the right, the dome with the telescope installed within.}
  \label{arcade}
 \end{figure}

Once the ARCADE project ended, the telescope was shipped back from Colorado to Italy for optimization and upgrade before its transfer 
to the first of the CTA sites. The upgrade of the hardware has been completed, and presently the Lidar 
is in L'Aquila, Italy to undergo quality tests 
before the shipment to the ORM, which is expected to happen before the end of 2017.
The ARCADE Lidar is a newtonian telescope (diameter of 20 cm, F3) with a UV Nd:YAG laser source (355 nm, maximum pulse
energy 5 mJ and variable repetition rate 1 - 100 Hz) designed to measure the elastic and Raman 
(nitrogen and water vapour) backscattered photons. 
In figure \ref{arcade} a picture of the telescope is shown. For a detailed description of the former ARCADE system, see \cite{arcade_icrc15}.

\begin{figure}[h!t]
   \centering
   \includegraphics[width=.5\textwidth]{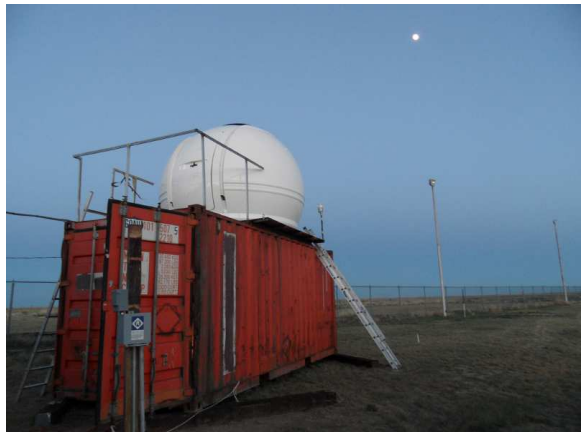}
  \caption{The astronomical dome containing the Lidar, on top of the 20 ft shipping container used in Lamar, Colorado}
  \label{dome}
 \end{figure}

The telescope is vertically steerable from 0 to 90 degrees and is hosted in an astronomical dome that is expected to be 
positioned on top of a 20ft shipping container, see \ref{dome}. 
The data acquisition system, dedicated PC and electronics will be hosted 
inside the shipping container.  The whole system only needs power and internet connection to run, and can be remotely controlled. 
A weather station provides information on rain, 
wind speed and temperature. The dome can be opened and closed remotely and operates smoothly also in difficult weather conditions.

\begin{figure}[h!t]
   \centering
   \includegraphics[width=.55\textwidth]{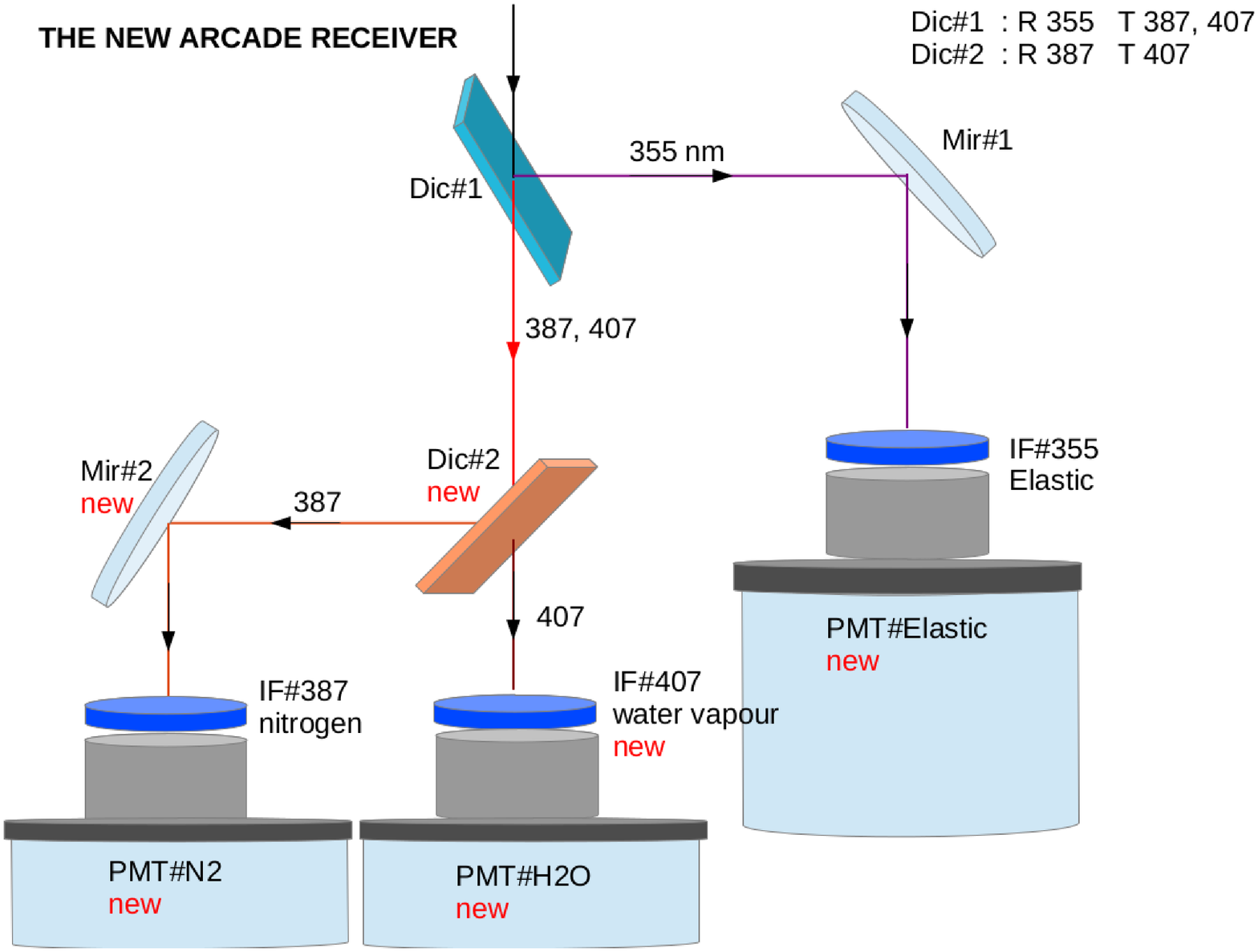}
   \includegraphics[width=.25\textwidth]{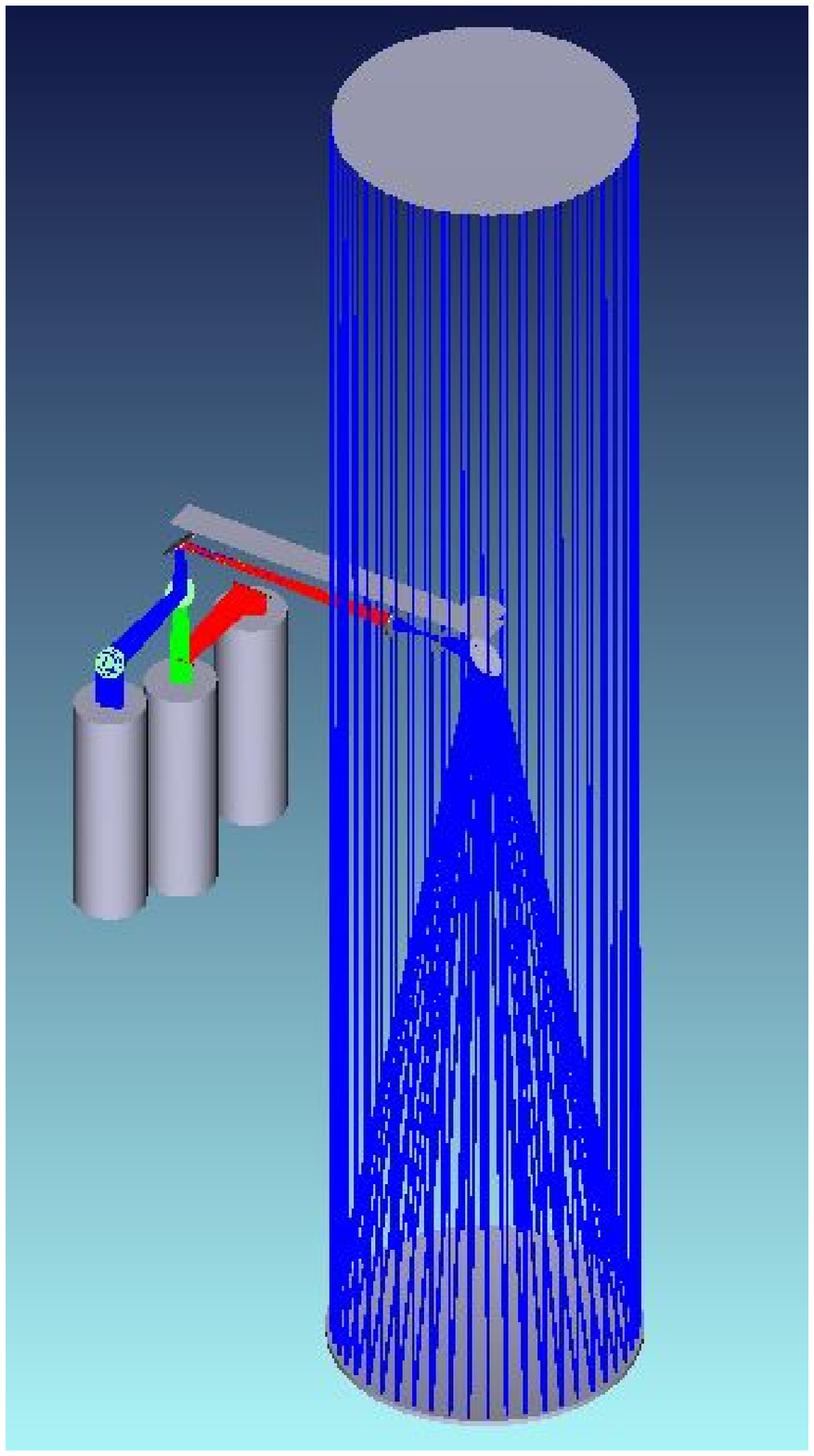}
  \caption{Left : a schematization of the new receiver. Right : Zemax simulation of
the Newtonian telescope, mirrors, beam splitters, interference filters and PMT photocatodes
for the detection of Rayleigh/Mie backscatter by air and aerosols, and N2 and H2O Raman backscatter. 
}
  \label{receiver}
 \end{figure}

The receiver has been upgraded at the INFN mechanical workshop of Torino, Italy to add a second Raman channel that
will improve the precision of the measurements and will allow to obtain the water vapour profiles of the atmosphere. Most
of the configuration of the receiver was preserved. A new dichroic mirror was added to separate the two Raman wavelenghts 
(407 nm for the water vapour and 387.5 for the nitrogen) before sending the light to the very narrow OD7 filters and photomultipliers. 
The design of the new receiver is shown in figure \ref{receiver}, together with a Zemax
simulation of the telescope, that allows to design the optical settings to achieve the best 
collecting efficiency of the lidar receiver, i.e., negligible optical overlap modulation.

The improvements to the system include: 

\begin{itemize}
\item{new Electron Tubes 9829B photomultipliers (the old ones were very old spares recovered from a previous experiment);}
\vspace{-0.3cm}
\item{new Isocomp APCv26 modules for the DAQ;}
\vspace{-0.3cm}
\item{addition of the water vapour Raman channel to the receiver. The water vapour profiles, measured simultaneously to 
the aerosol attenuation profiles, can help to define the physical properties of the aerosol particles;}
\vspace{-0.3cm}
\item{replacement of the damaged optical elements (primary mirror and some of the optics in the laser bench).} 
\end{itemize}

In L'Aquila, everything is now ready for the full assembly of the laser bench and the receiver, whose
parts are being verified separately. The characterization of the performance of the optical components and of the detectors 
(photomultipliers) are complete, 
while its documentation is in preparation. A software for the handling of the remote and automatic functionality of ARCADE is under
development, while the software packages to control the laser and data acquisition system are ready. During this month, the
first direct atmospheric measurements are scheduled and the system is expected to be shipped to La Palma before the end of 2017.

\section{Atmospheric Simulations} 

\begin{figure}[h!]
   \centering
   \includegraphics[width=.45\textwidth]{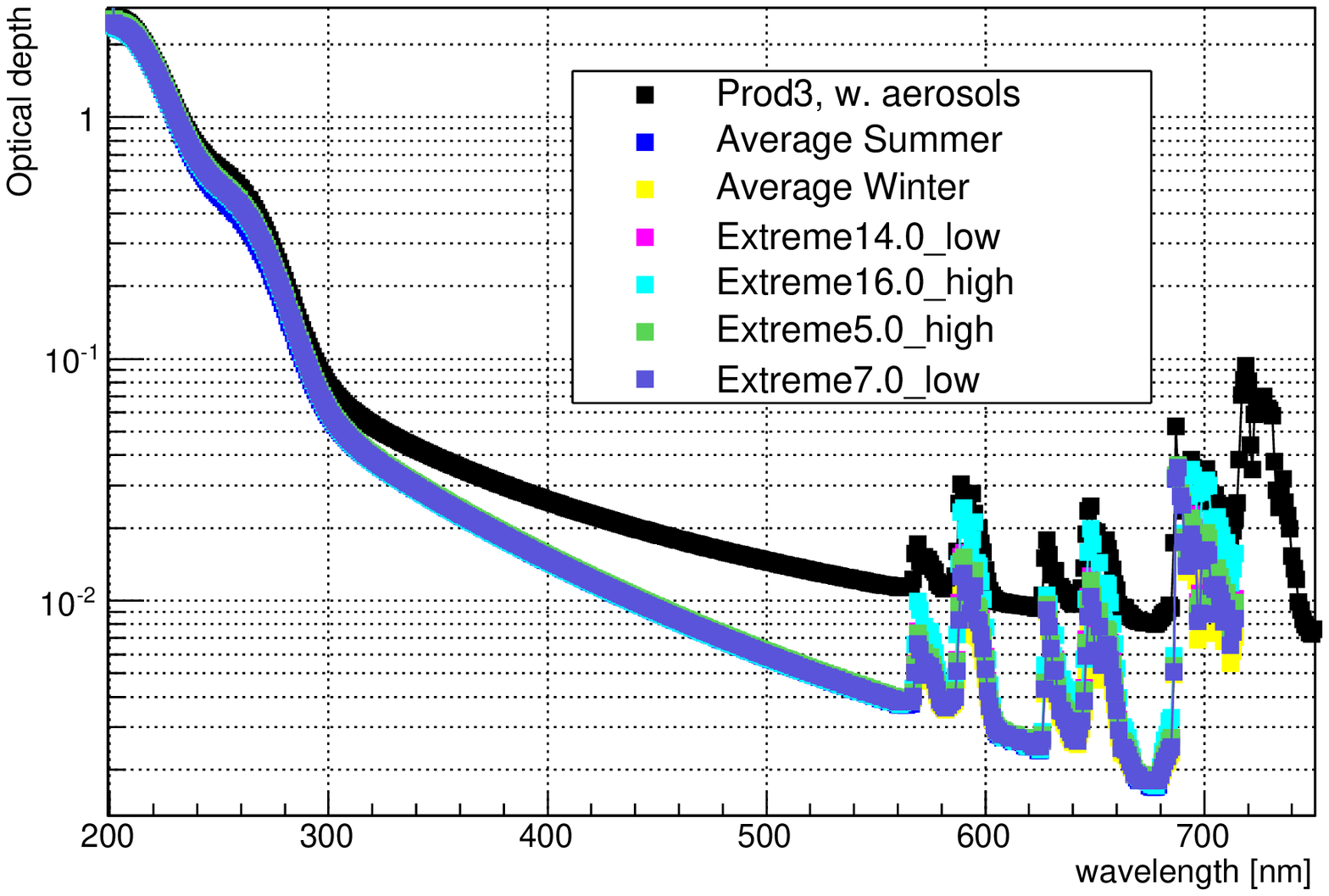}
   \includegraphics[width=.45\textwidth]{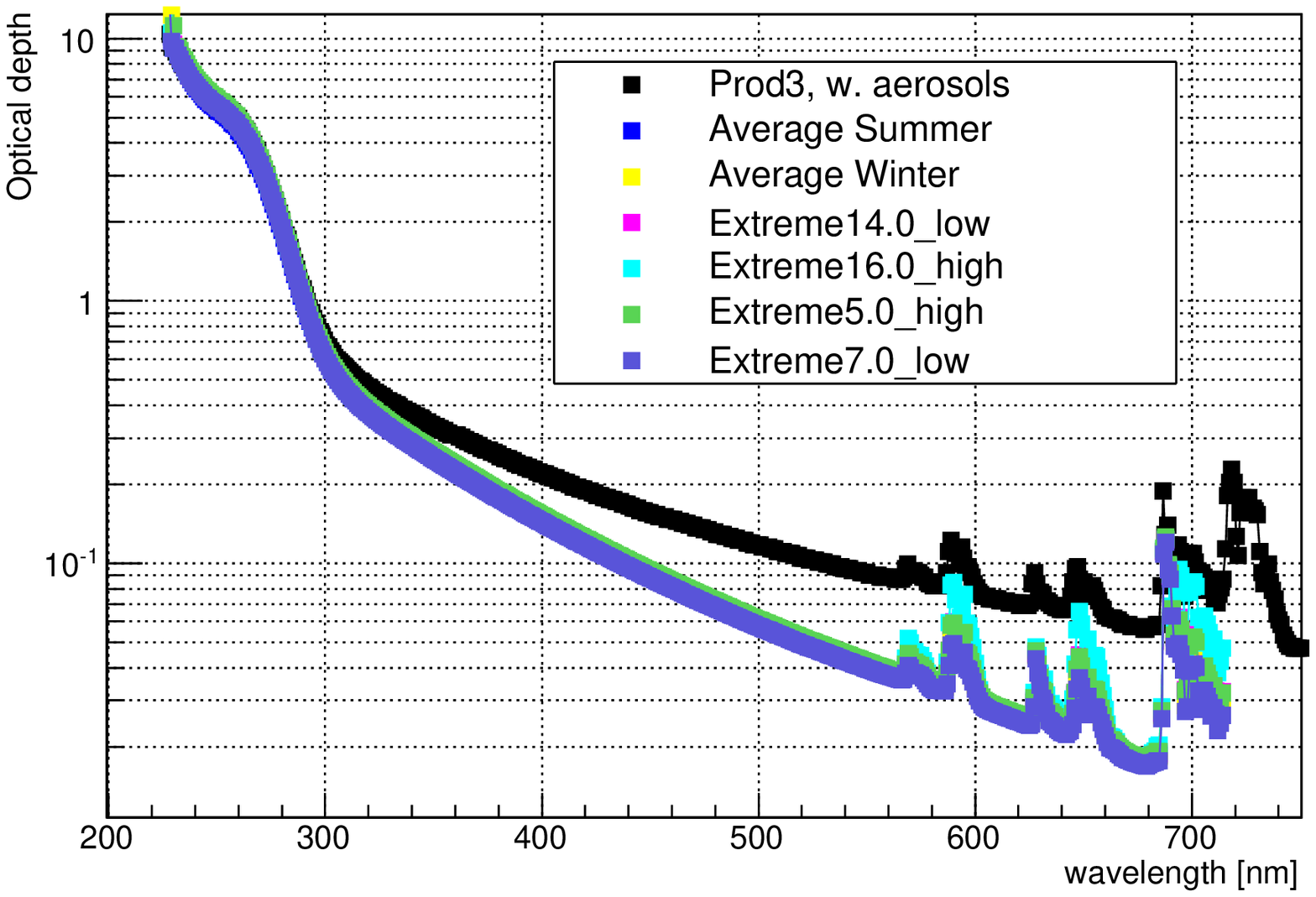}
  \caption{Optical depth profiles at 2.647 km (left) and 8 km (right)
   for the different molecular profiles (see text).}
  \label{sim_1}
 \end{figure}

A detailed Monte Carlo simulation, including the shower development in the atmosphere and the Cherenkov light production and the detector 
response, has been used to estimate the effect of different atmospheric profiles and aerosols (dust, clouds) 
on the CTA performances, as the effective area (energy threshold and flux), energy bias and resolution, and angular resolution. 
The shower development in the atmosphere has been simulated using the CORSIKA\cite{corsika} package, while for the detector response a specific 
software (Sim$\_$Telarray) has been developed within the CTA Consortium. One of the important points is to establish how  
the uncertainties related to atmospheric calibration affect the reconstructed energy and flux, 
and which is their compliance with the CTA performance requirements.
Studies on the atmospheric molecular profiles 
have already been carried out, focused on the La Palma site due to the well known atmosphere, with a lot of measurements already available. 
Using the GDAS profiles \cite{gdas} (measured up to 25 km) and NRLMSISE-00 \cite{nrmlsise} (from 25 to 100 km) and an exponential 
extrapolation up to 120 km, the following atmospheres have been selected:

\begin{itemize}
\item average winter, obtained averaging over all the winter profiles;
\vspace{-0.3cm}
\item average summer, obtained averaging over all the summer profiles;
\vspace{-0.3cm}
\item two extreme atmospheric profiles, with a minimum air density at 7 or 14 km;
\vspace{-0.3cm}
\item two extreme atmospheric profiles, with a maximum air density at 5 or 16 km.
\vspace{-0.3cm}
\item the standard CTA Monte Carlo production (Prod3, including aerosols)
\end{itemize}

where "extreme" is defined as the strongest deviation of the density from a mean model found at a given altitude in a period of 3 years of GDAS data.
The altitude of the extreme model has been chosen to maximize the effect on the low or on the high energy events, due to the fact that the 
low energy events develop high in the atmosphere, while the Cherenkov light contribution of high energy events comes mainly from low altitudes. 
The density profiles of all these models are indeed quite similar, with a maximum difference of $\pm$ 10\% with respect to the average winter 
model used as a reference. Also the thickness and refractive index differences are within 10\% or less. Fig. \ref{sim_1} shows the optical 
depth profiles at 2.647 and 8 km, obtained using MODTRAN with the previous profiles as input and molecular extinction only. 
All the obtained optical depths are very similar except for the Prod3 simulation \cite{prod3} for which the aerosols were included. 
From these plots it becomes clear that the biggest variations of molecular profiles measured on site have less impact, 
while aerosols do have a larger effect and must be continuously monitored.

\begin{figure}[h!]
   \centering
   \includegraphics[width=.45\textwidth]{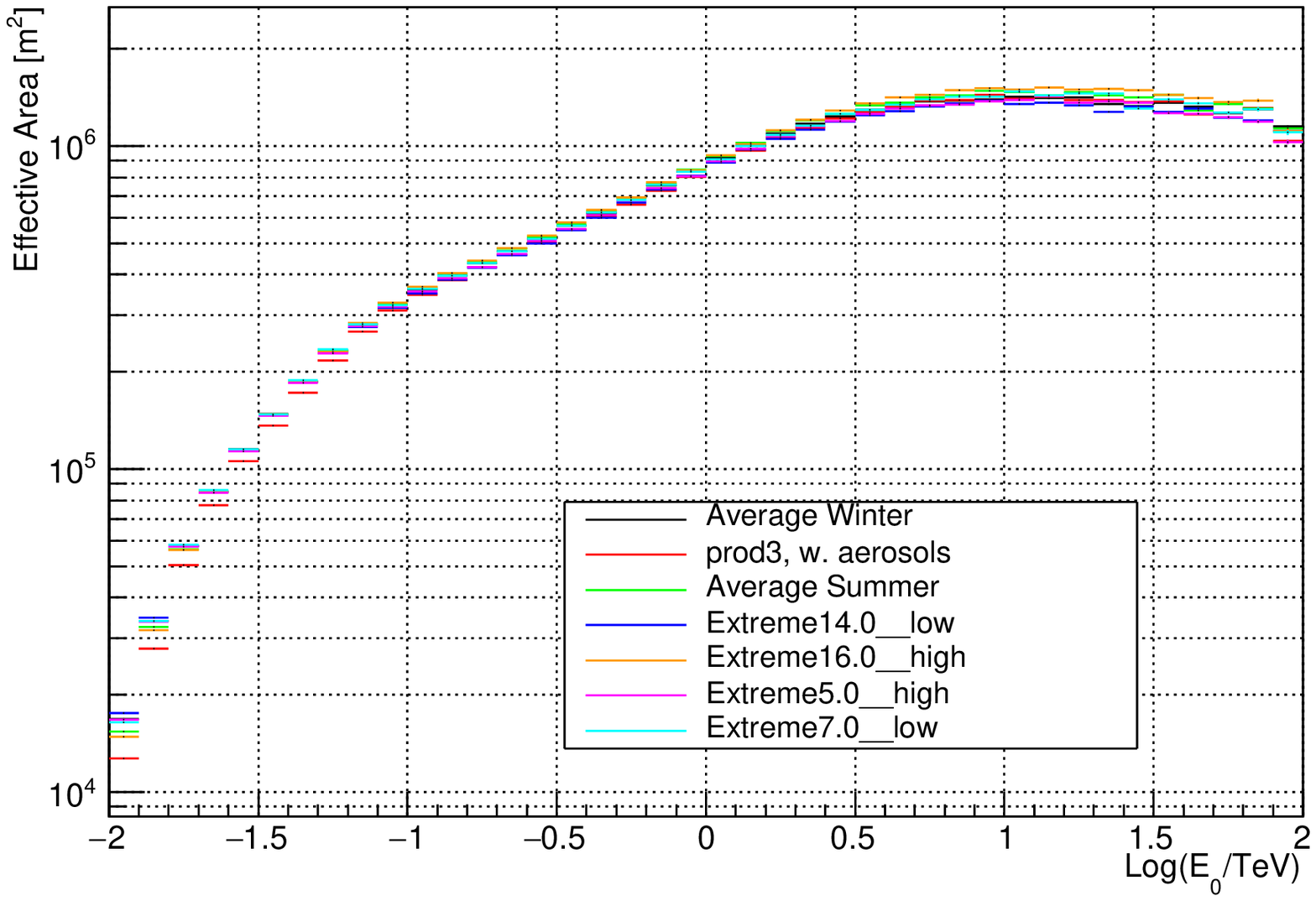}
   \includegraphics[width=.45\textwidth]{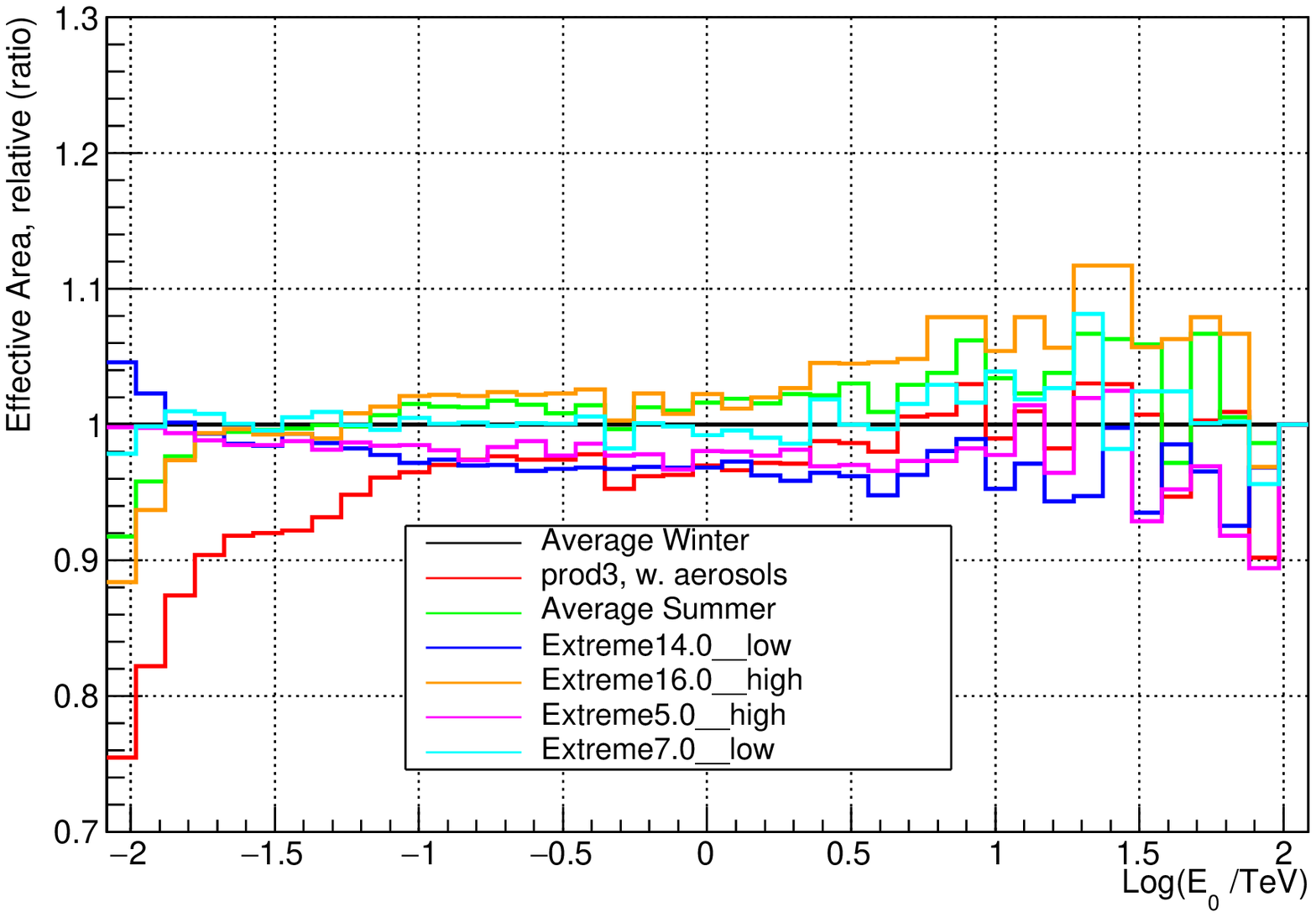}
  \caption{Effective area (left) for the different molecular profiles and
   the relative ratio with respect to Average Winter (see text).}
  \label{sim_2}
 \end{figure}

Nevertheless, to accurately estimate the 
effect of the molecular profile variations 
on the Cherenkov light detection, a Monte Carlo simulation as previously described was performed. For each molecular atmospheric profile, a set
 of primary gamma rays at a zenith angle of 20 degrees in the energy range 3 GeV - 100 TeV and spectral index $\alpha$ = -2 has 
been developed in the atmosphere and then detected by a possible CTA-North array made of 4 LSTs and 15 MSTs. 
The reconstruction has been made with the standard Prod3 analysis options and cuts, and since no background hadrons have been 
simulated, the obtained performances are overestimated, but can be compared to each other: an evaluation of this overestimation can be 
obtained comparing the results of this work with the Prod3 ones, for which the full analysis was carried out. Fig. \ref{sim_2} shows 
the effective area and its relative ratio with respect to the Average Winter case. As expected, the maximum difference is visible with respect to 
the atmosphere that includes aerosols. 

\begin{figure}[h!]
   \centering
   \includegraphics[width=.45\textwidth]{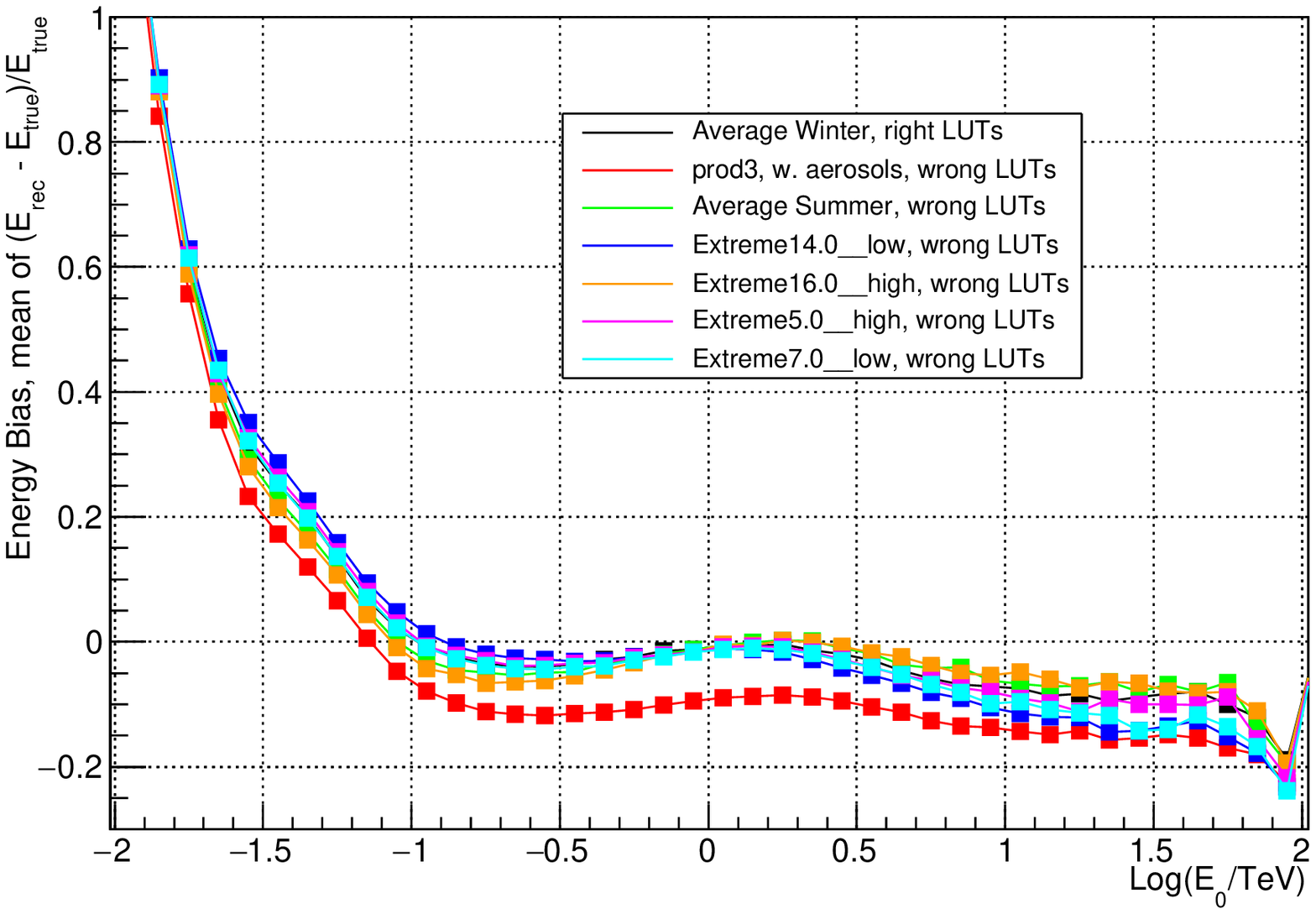}
   \includegraphics[width=.45\textwidth]{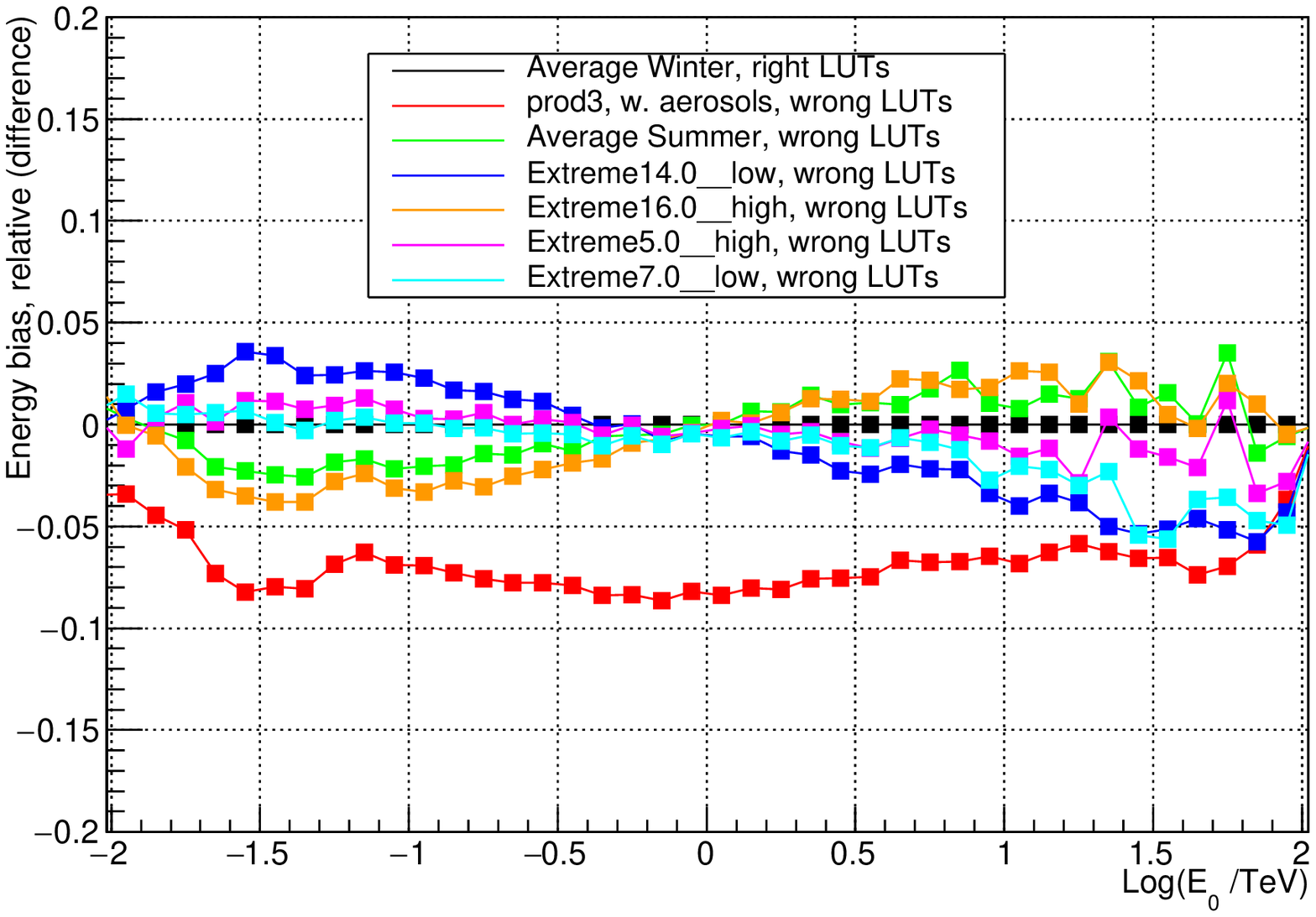}
  \caption{Energy bias (left) for the different molecular profiles and
   the relative difference compared to Average Winter (see text). The
   Average Winter LUTs have been used for all atmospheric models.}
  \label{sim_3}
 \end{figure}

Among the different molecular profiles, the maximum effect is for the extreme models, but never exceeding 10\%. At higher energies, 
a denser atmosphere at high altitudes (about 14 km) produces larger effective areas, while at lower energies a denser atmosphere 
at low altitudes (< 7 km) produces larger effective areas. The effect of different molecular profiles is even smaller on the energy bias, 
being at a level of $\pm$ 4\% above 20 GeV. The energy resolution is at the same level, while the differences in the angular resolution are within 0.05 degrees. 
More interesting is the effect of using the average winter LUTs to reconstruct data produced with all other atmospheric profiles. 
In this case we evaluate the necessary knowledge of the molecular profile testing the worst case, i.e. no information at all on the 
specific situation. We expect a large effect on the energy bias, since the same size corresponds to different energies, but only minor 
effects on the energy resolution, since the spread of the reconstructed energies should not increase. Fig. \ref{sim_3} shows the results 
in terms of energy bias and relative difference compared to the Average Winter model. The difference is largest when considering the aerosols 
(rather constant 8\% energy underestimation) while the purely molecular profiles shows an energy dependency within $\pm$ 4\%, with the largest 
effect for the extreme atmospheric profiles occurring at high altitudes. As expected, the energy and angular resolution are poorly affected 
($\pm$ 2\% and +0.02; -0.005 degrees respectively). 

The next steps will be to include aerosols of different nature and attenuation profiles in the simulation: the ARCADE measurements
on site will help to define the possible range for the aerosol stratification. We also plan to study, with the MODTRAN model, the dependencies 
of the Optical Depth on the variation of the atmospheric profiles (thickness, refractive index, RH, Water Vapour fraction, and so on) to understand 
the most important parameters for the allowed variability. Finally, we will study the effect of the uncertainties on the currently foreseen 
CTA measurements, answering the question of how precisely they need to be measured.

\section{Acknowledgments}
This work was conducted in the context of the CTA CCF Work Package. We gratefully acknowledge financial support from the agencies and organizations listed here : http://www.cta-observatory.org/consortium\_acknowledgments.


\begin{thebibliography}{99}
\bibitem{jan_icrc17} J. Ebr et al., \emph{Atmospheric calibration of the Cherenkov Telescope Array}, these proceedings.
\bibitem{raman_icrc17} Vasileiadis G. et al., \emph{Raman LIDARs for atmospheric calibration in CTA}, these proceedings.
\bibitem{arcade_icrc15} L. Valore et al.,  \emph{The ARCADE Project}, Proceedings of \emph{34th ICRC 2015}.
\bibitem{Jurysek} Jurysek, J. et al., \emph{Sun/Moon photometer for Cherenkov Telescope Array}, these proceedings.
\bibitem{raman_Auger} C. Medina et al.,  \emph{Studies in the atmospheric monitoring at the Pierre Auger Observatory using the upgraded Central Laser Facility}, Proceedings of \emph{34th ICRC 2015}.
\bibitem{corsika} Heck, D., Knapp, J., Capdevielle, J.N., Schatz, G., \& Thouw, T.   1998, Forschungszentrum Karlsruhe Report FZKA, 6019.
\bibitem{gdas} https://www.ncdc.noaa.gov/data-access/model-data/model-datasets/global-data-assimilation-system-gdas
\bibitem{nrmlsise} https://ccmc.gsfc.nasa.gov/modelweb/atmos/nrlmsise00.html
\bibitem{prod3} G. Maier et al., \emph{Performance of CTA}, these proceedings.
\end{thebibliography}
\end{document}